\documentclass[journal]{IEEEtran}

\usepackage{cite}
\usepackage{amsmath,amssymb,amsfonts}
\usepackage{algorithmic}
\usepackage{graphicx}
\usepackage{textcomp}
\usepackage{xcolor}
\usepackage{footnote}
\usepackage{hyperref}
\usepackage{xcolor}
\usepackage{tabu}
\usepackage{makecell}
\usepackage{float}
\usepackage{adjustbox}
\usepackage[linesnumbered,ruled]{algorithm2e}
\usepackage{multirow,booktabs}
\usepackage[inline]{enumitem}
\usepackage{lipsum}
\usepackage{dblfloatfix}
\usepackage[]{caption}
\usepackage{subcaption}
\usepackage{soul}
\usepackage{multirow}
\usepackage{comment}
\begin{document}
\title{Dynamic Traffic Allocation in IEEE 802.11be Multi-link WLANs} 
\author{Álvaro~López-Raventós, and Boris~Bellalta% <-this % stops a space
\thanks{All the authors are with the \textit{Wireless Networking (WN)} research group at \textit{Universitat Pompeu Fabra, Barcelona, Spain} (e-mail: {alvaro.lopez, boris.bellalta}@upf.edu). This  work  has  been  partially  supported by the Spanish Government under grant WINDMAL PGC2018-099959-B-I00 (MCIU/AEI/FEDER,UE), and Cisco.}}%% <-this % stops a space
\maketitle

\begin{abstract}
The multi-link operation (MLO) is a key feature of the next IEEE 802.11be Extremely High Throughput amendment. Through its adoption, it is expected to enhance users' experience by improving throughput rates and latency. However, potential MLO gains are tied to how traffic is distributed across the multiple radio interfaces. In this paper, we introduce a traffic manager atop MLO, and evaluate different high-level traffic-to-link allocation policies to distribute incoming traffic over the set of enabled interfaces. Following a flow-level approach, we compare both non-dynamic and dynamic traffic balancing policy types. The results show that the use of a dynamic policy, along with MLO, allows to significantly reduce the congestion suffered by traffic flows, enhancing the traffic delivery in all the evaluated scenarios, and in particular improving the quality of service received by video flows. Moreover, we show that the adoption of MLO in future Wi-Fi networks improves also the coexistence with non-MLO networks, which results in performance gains for both MLO and non-MLO networks.
\end{abstract}


% ---------------

\section{Introduction}

The multi-link operation (MLO) is a revolutionary feature that is planed to be part of the IEEE 802.11be Extremely High Throughput (EHT) amendment~\cite{draft11be}. By the use of multiple radio interfaces, MLO-capable devices will be able to send and receive traffic over different wireless links, allowing devices to experience higher throughput rates, as well as lower end-to-end latency delays. To support such implementation, the Task Group~"be"~(TGbe) has proposed several modifications to the standard, being the nodes' architecture one of the most significant. In this regard, it is suggested to split common and link-specific medium access control (MAC) functionalities into two different levels~\cite{lopez2022multi}.

With such approach, the TGbe aims to provide nodes with a dynamic, flexible, and seamless inter-band operation. To that end, a unique MAC instance is presented to the upper-layers, while each interface is able to maintain an independent set of channel access parameters~\cite{LevyEHT2020}. However, proper traffic balancing over the different interfaces is required to make the most out of the MLO. To implement such a load balancing, we rely on the existence of a traffic manager on top of the MLO framework, in order to apply different traffic management policies to allocate new incoming flows/packets across the enabled interfaces\footnote{We refer to enabled interfaces as those in which at least a wireless link has been established.}. This approach allows to control the allocation process, ensuring a more balanced usage of the network resources.

Although MLO is gaining relevance at a very fast pace, none of the existing works have tackled how traffic allocations may be performed. For instance, existing MLO works relate to feature improvements, as the work in~\cite{naik2021can}, in which the authors prove that MLO can reduce latency by means of minimizing the congestion. Similarly, \cite{carrascosa2021experimental} shows experimentally that MLO is able to reduce Wi-Fi latency in one order of magnitude in certain conditions by just using two radio interfaces. Additionally, authors in~\cite{yang2019ap} suggest that the use of MLO per-se may not be sufficient enough to provide the prospected gains without a coordination between access points (AP) in high density areas. Hence, they propose a coordination framework to achieve high throughput in those circumstances. On the other hand, works in \cite{naribole2020simultaneous, naribole2020sim} focus on maximizing the medium utilization, while the interference suffered by constrained nodes is minimized. As shown, none have tackled neither the implementation of a traffic manager atop MLO, nor considered the performance gains from a flow-level perspective.

A first evaluation of the capabilities of the proposed traffic manager was presented in~\cite{lopez2021ieee}. There, it was shown ---as expected--- that congestion-aware policies outperform a blindfolded scheme. Additionally, and more important, it was shown that allocating the whole traffic of an incoming flow to the emptiest interface was almost as good, as proportionally distributing the flow over multiple interfaces. Such finding relies on the fact that using more interfaces, a traffic flow becomes more vulnerable to suffer a congestion episode due to the changing spectrum occupancy conditions caused by the neighboring wireless local area networks (WLANs).

In this letter, we introduce and evaluate a dynamic traffic balancing policy for the traffic manager, which periodically modifies the traffic-to-link allocation accordingly to the instantaneous channel occupancy conditions. Thus, we expect to minimize the negative impact of neighboring WLANs over the traffic flows by reacting to changes in the spectrum occupancy. The presented results show that the application of a dynamic policy has a significant impact on the spectrum usage efficiency, while improving the service received by the flows. For instance, we observe that video flows are able to keep up to 95\% their performance in most of the scenarios, when the dynamic policy is applied. Additionally, we showcase that the adoption of MLO in future Wi-Fi networks eases coexistence issues with non-MLO networks, which performance is improved up to 40\% when surrounded by MLO~BSSs.

% ---------------

\section{Policy-based traffic management for MLO-capable WLANs}\label{sec:policies}

\begin{figure}[t]
    \centering
    \includegraphics[width=\linewidth]{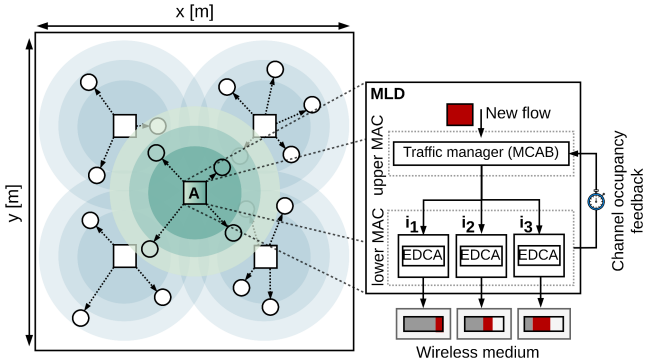}
    \caption{Scenario and architecture representation. The high, medium and low shaded areas represent the operation range for the 6~GHz, 5~GHz and 2.4~GHz bands, respectively. Colored in red is represented the traffic allocated to each interface, whereas in gray is represented the channel occupancy.}
    \label{fig:scenario}
\end{figure}

The multi-interface availability allows to naturally think of a manager in order to distribute traffic. Following the proposals of the TGbe, this logical entity should be placed at the upper MAC level, since the interface assignation is performed once traffic goes through it~\cite{Hamilton2021}. Once a connection\footnote{A connection is defined to be a logical link, in which traffic of a certain application is exchanged between two end hosts.} is established between an AP-STA pair, and traffic streams start to flow, the traffic manager is in charge to allocate the traffic to the corresponding interfaces. Such approach allows to not only achieve an efficient use of the network resources, but better control the capabilities of multi-link devices (MLDs) supporting, for instance, advanced traffic differentiation, beyond the default MLO's TID-to-link mapping functionality~\cite{lopez2022multi}. Figure~\ref{fig:scenario} shows an schematic of a MLD architecture, with a traffic manager~representation.

To perform the allocation process, the transmitting MLD gathers the instantaneous channel occupancy at each interface according to the set of enabled interfaces at the receiving node. Then, the traffic manager is able to ensure that the transmitting MLD will not allocate traffic to congested interfaces, distributing it over all of them proportionally to their occupancy. At the following, we present the different policies, which can be classified into non-dynamic and dynamic in regards of their behavior.

\subsection{Non-dynamic congestion-aware policies}
Under a non-dynamic strategy, each flow maintains the same traffic-to-link allocation during its lifetime. That is, upon a flow arrival, the channel occupancy is gathered, and the traffic is distributed either proportionally over multiple interfaces according to their congestion, or fully into the less congested one. We define two different non-dynamic policies:

\begin{itemize}
    \item \textbf{Single Link Less Congested Interface (SLCI).} Upon a flow arrival, pick the less congested interface, and allocate the new incoming flow to it.
    \item \textbf{Multi Link Congestion-aware Load balancing at flow arrivals (MCAA).} Upon a flow arrival, distribute the new incoming flow's traffic accordingly to the observed channel occupancy at the AP, considering the enabled interfaces of the receiving station. Namely, let $\rho_i$ the percentage of available (free) channel airtime at interface~$i$. Then, the fraction of the flow's traffic allocated to interface~$i$ is given by $\ell_{i \in \mathcal{J}}~=~\ell \frac{\rho_i}{\sum_{ \forall j \in \mathcal{J}}{\rho_j}}$, with $\ell$ being the traffic load, and $\mathcal{J}$ the set of enabled interfaces at the target station. If there are any other active flows at the AP, their traffic allocation remain the same as it was.
\end{itemize}

Due to their straightforward approach, the application of non-dynamic policies are well-suited for scenarios where the interfaces' congestion levels remains almost stationary. Their computational cost is low, as only few calculations are done at flow arrivals. 

\subsection{Dynamic congestion-aware policies}
A dynamic strategy is able to periodically adjust the traffic-to-link allocation in order to follow channel occupancy changes, and so, taking the most out of the different enabled interfaces. In this regard, a traffic (re)allocation may be triggered by two different events: a new flow arrival or a periodic timer, which wakes up every~$\delta$ units of time. Under both events, the channel occupancy is gathered to proportionally (re)distribute the traffic load of all active flows to any of the enabled interfaces. It is worth mention that, the dynamic reallocation of traffic is performed by adjusting the interfaces' traffic weights (i.e., traffic percentage associated to each one), which are tracked by the traffic manager at the upper MAC level. Besides, we consider such reallocation to be instantaneous. We define the following dynamic policy:

\begin{itemize}
    \item \textbf{Multi Link Congestion-aware Load balancing (MCAB).} Upon a flow arrival or at every $\delta$ units of time, collect the channel occupancy values and sort all flows (including the incoming one) in ascending order, considering the number of enabled interfaces at the destination station (i.e., first the flows with less enabled interfaces). In case two or more flows have the same number of enabled interfaces in the destination station, they are ordered by arrival time. After, start (re)allocating the flows' traffic accordingly to the same procedure as in MCAA. 
\end{itemize}

Through its dynamic implementation, the MCAB minimizes the effect of neighboring BSSs actions, as they usually result in abrupt changes in the observed congestion at each link. Therefore, such policy scheme is able to adjust the traffic allocated to each link, exploiting the different traffic activity patterns while maximizing the traffic delivery. However, it is noticeable that the MCAB gain is conditioned to perform multiple operations in shorts amounts of time, which may be impractical in high density areas, as the computational requirements to (re)distribute all flows grows with the number of active users.

% ---------------

\section{System model}

\subsection{Scenario}

To assess the performance of the different policies, we consider an scenario with $N$~BSSs, each composed by an AP and $M$~stations as depicted in Figure~\ref{fig:scenario}. In every scenario, we place the $\text{BSS}_\text{A}$ at the center, and the other $N-1$~BSSs are distributed uniformly at random over the area of interest. To consider a random generated scenario as valid, the inter-AP distance must be equal or higher than 3~m. Otherwise, the scenario is discarded and a new one is generated. For each BSS, stations are placed within a distance $d\in[1-5]$~m and an angle $\theta\in[0-2\pi]$ from its serving AP, both selected uniformly at random.

\begin{figure}[b!]
    \centering
    \includegraphics[width=.85\linewidth]{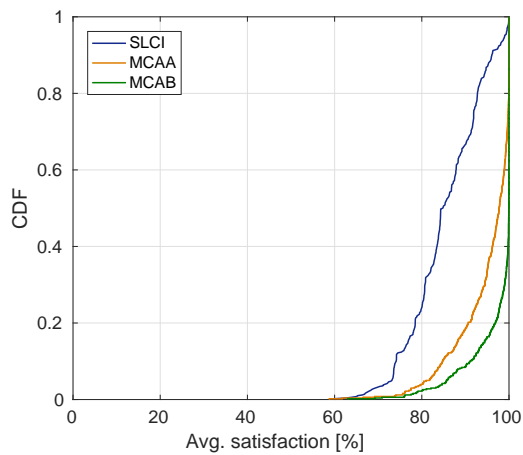}
    \caption{$\text{AP}_\text{A}$'s avg. satisfaction per policy.}
    \label{fig:cdf_sat}
\end{figure}

\subsection{Node operation}

All APs and stations have three wireless interfaces, each one operating at a different frequency band (i.e., 2.4~GHz, 5~GHz and 6~GHz). For each station, the set of enabled interfaces includes all the interfaces that can be effectively used (i.e., the power received from the serving AP is above the clear channel assessment (CCA) threshold). The modulation and coding scheme (MCS) used by the serving AP at each interface is selected accordingly to the signal-to-noise ratio (SNR). All stations are inside the coverage area of its serving AP for at least the 2.4~GHz band. All APs' interfaces corresponding to the same band are configured with the same radio channel.

Unless otherwise stated, all the APs and stations will be considered MLO-capable, using an asynchronous transmission mode~\cite{lopez2022multi}. Besides, except for $\text{AP}_\text{A}$, which will be set either with the SLCI, MCAA or MCAB, the rest of the APs will implement either the SLCI or MCAA policy schemes, which will be selected with the same probability. Regarding the MCAB policy, we set the time between two adaptation periods to be~$\delta$~s. In this paper, $\delta$ is set to 1 s. The MCAB dependency in regards of $\delta$ is kept out of this article due to space limitations.

\subsection{Traffic considerations}

Only downlink traffic is considered. The deployed stations are defined as data or video depending on the traffic that they will request. Also, only one connection is considered per station, which is set to be alive during the whole simulation time. Video traffic is modeled as a single Constant Bit Ratio (CBR) traffic flow of $\ell_S$~Mbps, whereas data traffic behaves following an ON/OFF Markovian model, where each ON period is treated as a new flow. Therefore, for data flows, their traffic load is $\ell_E$~Mbps during the ON period, and zero otherwise. Both ON and OFF periods are exponentially distributed with mean duration T$_{\text{ON}}$ and T$_{\text{OFF}}$, respectively.

\begin{table}[t!]
   \centering % center the table
   \caption{Evaluation setup}
   \resizebox{\linewidth}{!}{
    \begin{tabular}{ll} % alignment of each column data
       \toprule
       \textbf{Parameter} & \textbf{Description}\\ 
       \midrule
       Carrier frequency & 2.437 GHz/5.230 GHz/6.295 GHz\\
       Channel bandwidth & 20 MHz/40 MHz/80 MHz\\
       AP/STA TX power & 20/15 dBm\\
       Antenna TX/RX gain & 0 dB\\
       CCA threshold & -82 dBm\\
       AP/STA noise figure & 7 dB\\
       \makecell[l]{Single user \\ spatial streams} & 2\\
       MPDU payload size & 1500 bytes\\
       Path loss & Same as \cite{lopez2020concurrent}\\
       Avg. data flow duration & T$_{\text{on}}$ = 3 s\\
       \makecell[l]{Avg. data flow \\ interarrival time}  & T$_{\text{off}}$ = 1 s\\
       MCAB adaptation period & $\delta=$ 1 s \\
       Packet error rate & 10\%\\
       Simulation time & 120 s (1 simulation)\\
       Number of simulations & N$_{\text{s}}$ (variable)\\
       \bottomrule
       \end{tabular}}
       \label{table:params}
\end{table}

% ---------------

\begin{figure*}[t!!!]
\centering
\hspace*{\fill}%
 \begin{subfigure}[]{\linewidth}
    \centering
     \includegraphics[width=.91\linewidth]{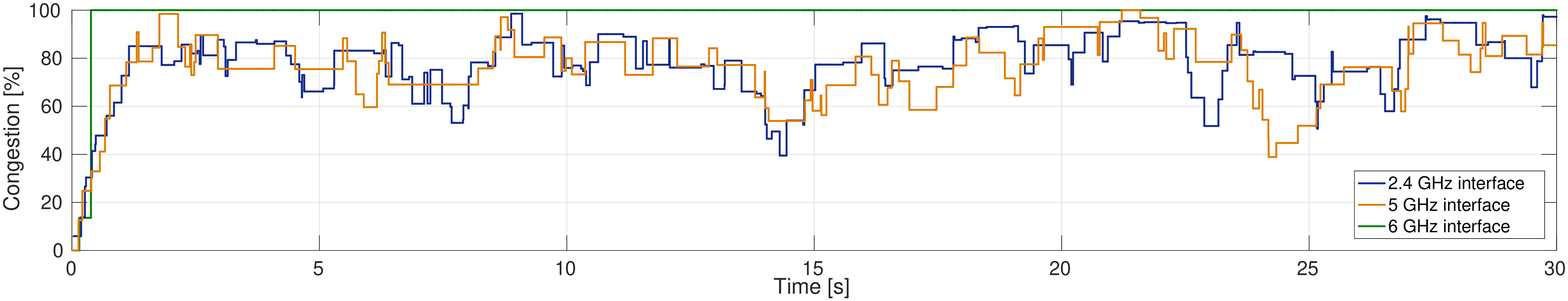}
     \caption{SCLI}
     \label{fig:cong_SLCI}
 \end{subfigure}
 \hspace*{\fill}%
 \begin{subfigure}[]{\linewidth}
 \centering
     \includegraphics[width=.91\linewidth]{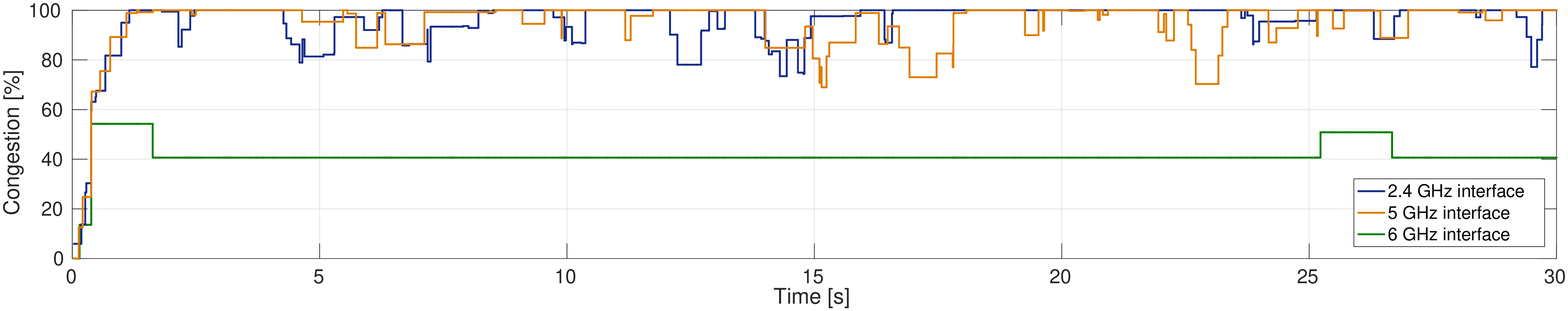}
     \caption{MCAA}
     \label{fig:cong_MCAA}
 \end{subfigure}
 \hspace*{\fill}%
  \begin{subfigure}[]{\linewidth}
 \centering
     \includegraphics[width=.91\linewidth]{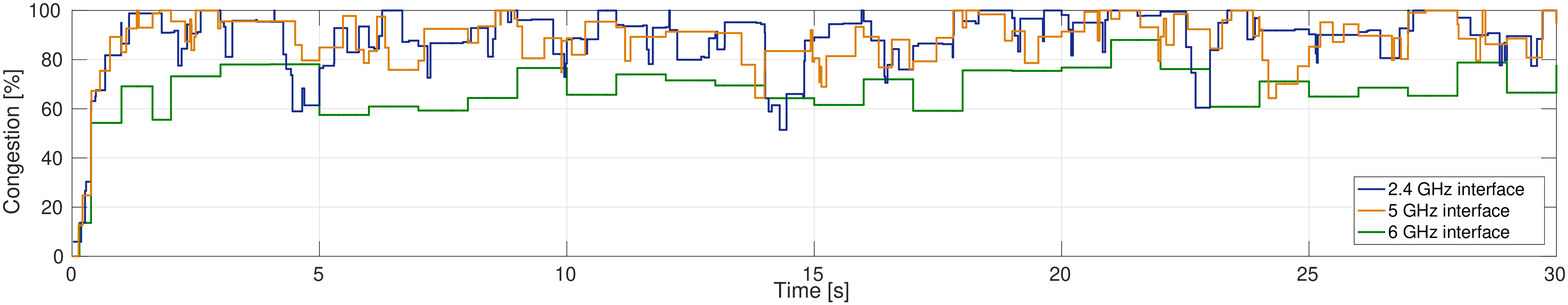}
     \caption{MCAB}
     \label{fig:cong_MCAB}
 \end{subfigure}
 \hspace*{\fill}%
\caption{$\text{AP}_\text{A}$'s congestion distribution per interface, and per policy application.}
\label{fig:congestion}
\end{figure*}

\section{Performance evaluation}

Flow-level simulations are performed using the Neko\footnote{The Neko simulation platform can be found in GitHub at: \url{https://github.com/wn-upf/Neko}} simulation platform, which implements the CSMA/CA abstraction presented in \cite{lopez2020concurrent}. This abstraction relies on the channel occupancy observed by each AP to calculate the allocable airtime for each flow, preserving the inherent Wi-Fi 'fair' share of the spectrum resources. Table~\ref{table:params} describes the complete set of parameters.

\subsection{Long-lasting flows}\label{subsec:longlast}

Here, we analyze the effects between the dynamic and non-dynamic traffic-to-link allocation policies in regards of video flows (i.e., flows with constant traffic requirements, and long lifetimes). To do so, we generate N$_{\text{s}}$~=~500 scenarios, placing $N$~=~5~BSSs over a 20x20 $\text{m}^{2}$ area. At the central BSS (i.e., $\text{BSS}_\text{A}$), we configure a unique video station with $\ell_S\sim\textit{U}[20,25]$~Mbps, whereas the remaining BSSs will have $M~\sim~\textit{U}[5,15]$ stations requesting data traffic with $\ell_E\sim\textit{U}[1,3]$~Mbps. 

Figure~\ref{fig:cdf_sat} plots the cumulative distribution function (CDF) of the average satisfaction ($\overline{s}$) experienced by the traffic flow served by $\text{AP}_\text{A}$, per policy type. We define $\overline{s}$ as the sum of the satisfaction of each station divided by the total number of stations in the BSS. Also, we refer to the satisfaction of a flow as the ratio between the allocated airtime by the AP during the flow lifetime, and the total amount of airtime required. As expected, the MCAB outperforms both non-dynamic policies. For instance, it is able to increase by 17\% and 6\% the $\overline{s}$ in regards of the MCAA and SLCI, respectively, for the 5\% worst case scenarios. Besides, we observe that the MCAB provides satisfaction values up to 95\% in more than the 90\% of the scenarios. This performance gains are provided by the periodic evaluation of the channel occupancy, which allows to leverage the emptiest interfaces, and so, making a better use of the available resources.

Further details are presented in Figure~\ref{fig:congestion}. There, we observe in detail the congestion evolution for each $\text{AP}_\text{A}$'s interface, during the first 30~s of a single simulation. Figure~\ref{fig:cong_SLCI} and Figure~\ref{fig:cong_MCAA} expose the main drawbacks of SLCI and MCAA, respectively, as the temporal evolution of the congestion reveals how unbalanced the interfaces are. First, the SLCI overloads the 6~GHz link by placing the whole video flow in it, while there is still room for some traffic in the other interfaces. On the contrary, the MCAA does not leverage the fact of having empty space at the 6~GHz interface, which makes the proportional parts of the flow allocated to the 2.4~GHz and 5~GHz links to suffer from congestion. Such inefficient operation from the non-dynamic policies is shown in Figure~\ref{fig:cong_MCAB} to be overcomed by the MCAB, as it reveals a more balanced use of the interfaces. However, we also observe that most of the time the congestion values for the 6~GHz interface are lower than for the other two. Such effect is related to the unequal number of neighboring nodes detected at each band. As a result, even if most of the traffic is allocated to this interface, it still manages to provide traffic with fewer congestion episodes. 

\subsection{Coexistence with legacy networks}

Wi-Fi's constant evolution makes newer devices, which implement up-to-date specifications, to coexist with others with less capabilities. As a result, last generation devices may decay in performance due to its coexistence with legacy ones. To assess if Multi-Band Single Link (MB-SL) BSSs affects the performance of MLO ones, we analyze four different cases in which we increment the fraction of MLO BSSs around the central one from 0, to 0.3, 0.7, and 1. To do so, we generate N$_{\text{s}}$~=~200 scenarios, placing $N$~=~11~BSSs. At the central BSS (i.e.,$\text{BSS}_\text{A}$), we configure a single video station with $\ell_S\sim\textit{U}[20,25]$~Mbps, whereas the remaining BSSs will have $M~\sim~\textit{U}[5,15]$ stations requesting background data traffic of $\ell_E\sim\textit{U}[1,3]$~Mbps. It is worth mention that, MB-SL APs are equipped with 3 interfaces, considering the associated stations are distributed across all three bands uniformly at random. 

\begin{figure}[t!!!]
\centering
\begin{subfigure}[]{.493\columnwidth}
    \centering
     \includegraphics[width=\linewidth]{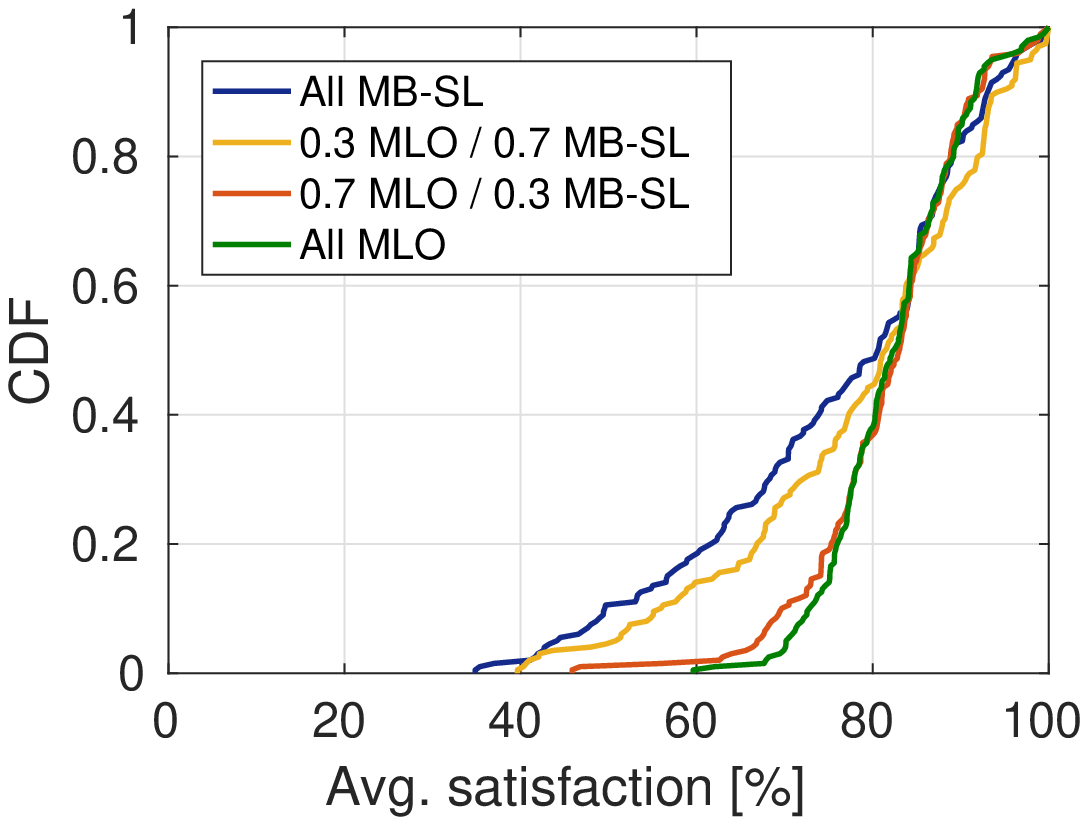}
     \caption{SCLI}
     \label{fig:Coex_SLCI}
\end{subfigure}
\begin{subfigure}[]{.493\columnwidth}
 \centering
     \includegraphics[width=\linewidth]{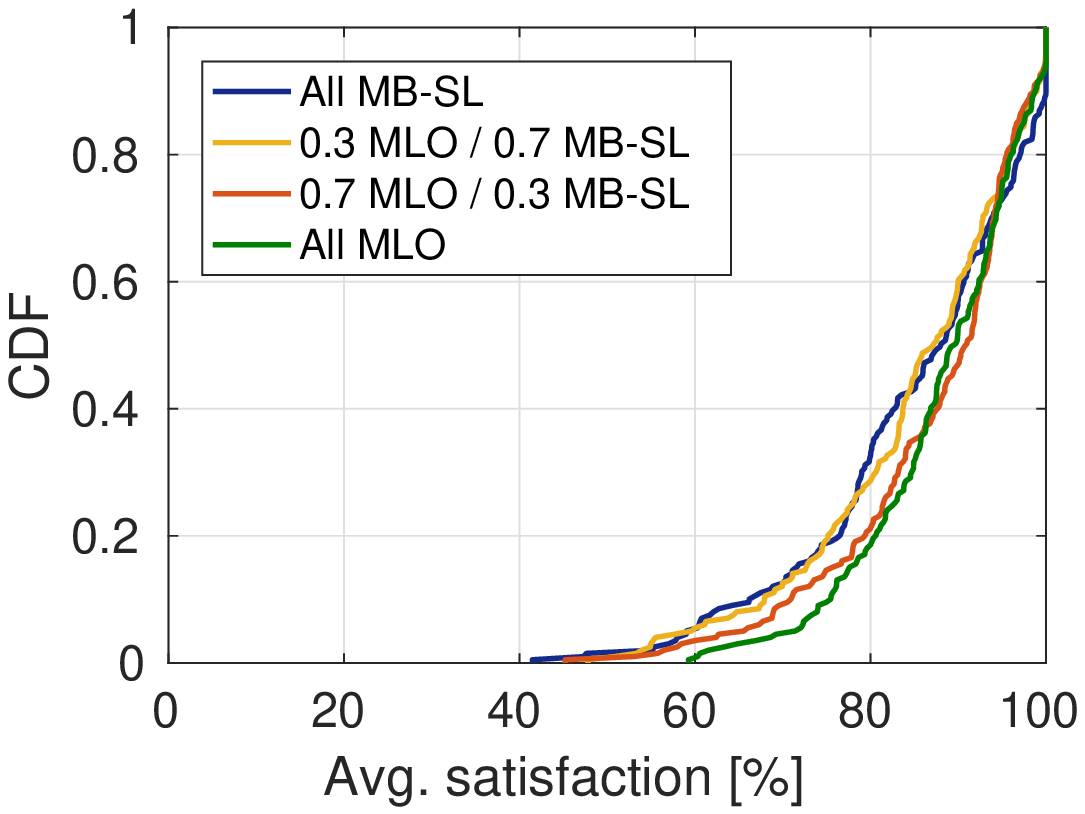}
     \caption{MCAA}
     \label{fig:Coex_MCAA}
\end{subfigure}
\begin{subfigure}[]{.493\columnwidth}
\quad
 \centering
     \includegraphics[width=\linewidth]{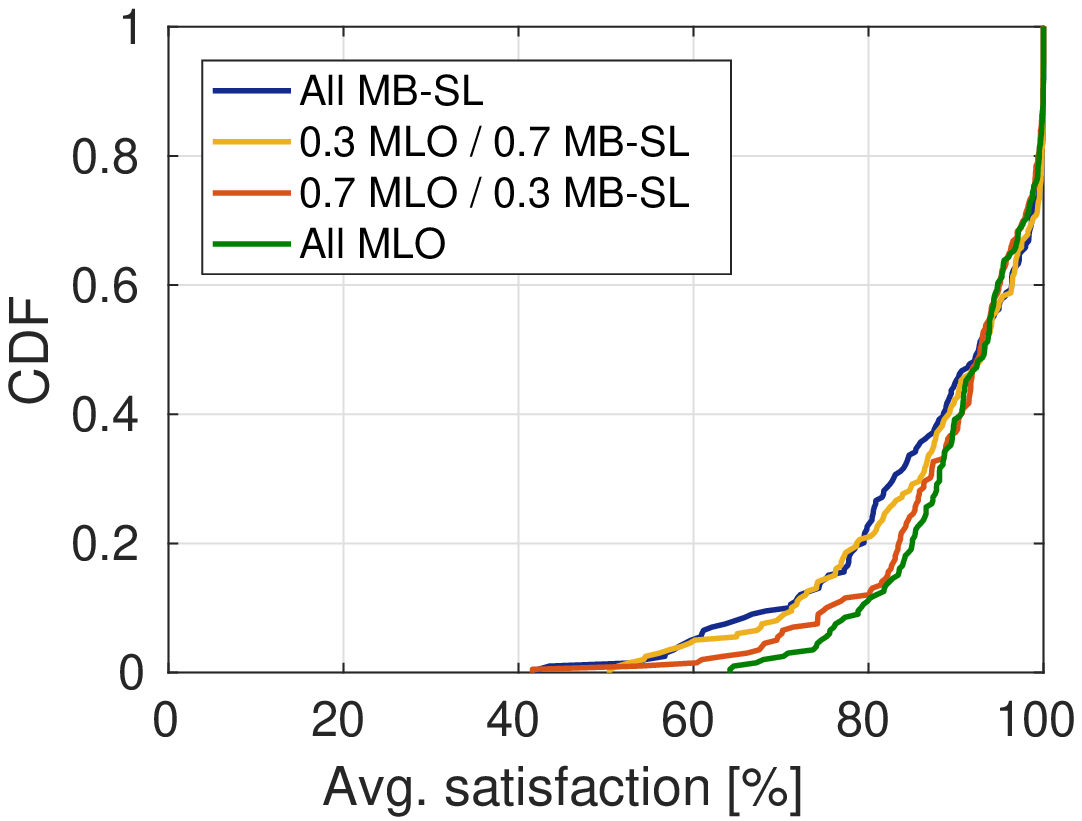}
     \caption{MCAB}
     \label{fig:Coex_MCAB}
\end{subfigure}
\begin{subfigure}[]{.493\columnwidth}
 \centering
     \includegraphics[width=\linewidth]{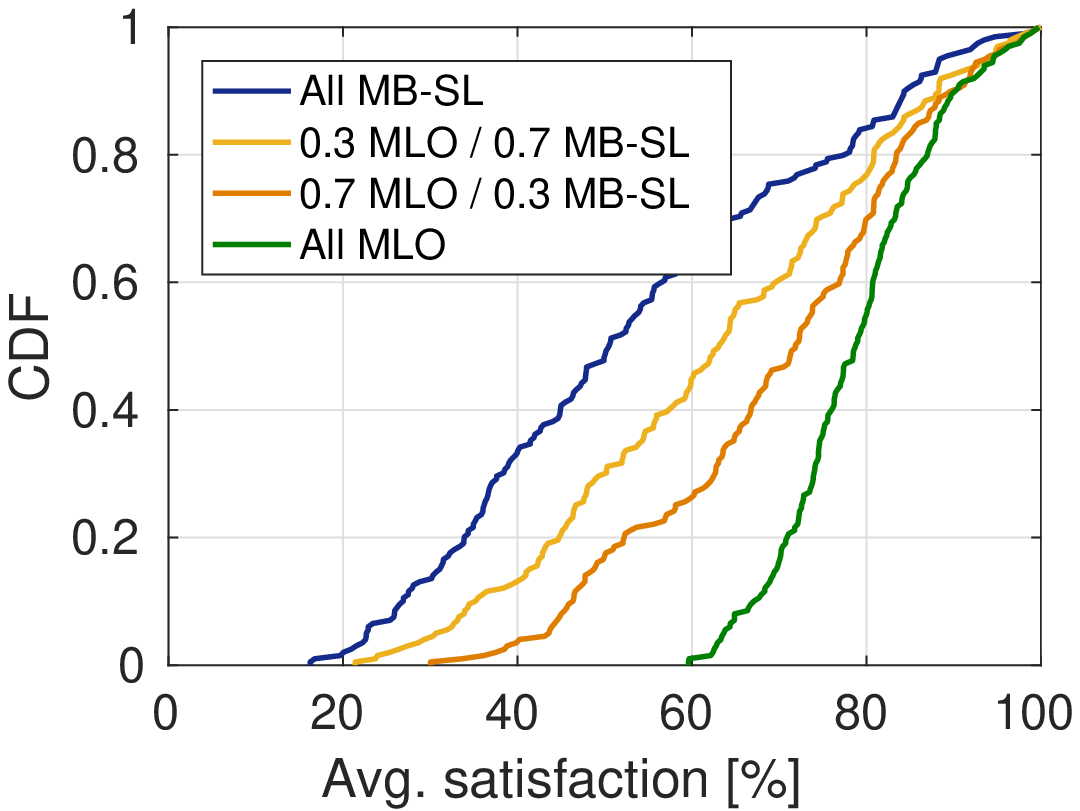}
     \caption{MB-SL}
     \label{fig:Coex_MBSL}
\end{subfigure}
\caption{Coexistence performance per policy type, and MB-SL.}
\label{fig:coexistence}
\end{figure}

Figures~\ref{fig:Coex_SLCI},~\ref{fig:Coex_MCAA}, and~\ref{fig:Coex_MCAB} show the CDF of the $\overline{s}$ for each policy. Regardless of the policy used, the central $\text{BSS}_\text{A}$ experiences a negative trend when it is surrounded by more legacy~BSSs, as the results show lower satisfaction values when so. Although the MCAA and MCAB experience low gains when increasing the number of MLO~BSSs, the SLCI presents a 17\% improvement for the 25th percentile, when comparing the performance results between the best and the worst (i.e., all MLO and all MB-SL, respectively) cases. Such an improvement is caused by the higher link availability from the neighboring BSSs to allocate traffic, which also avoid to overload the interfaces by the use of congestion-aware policies. On the other hand, comparing policies, we find that the MCAB outperforms the other ones. Specially, we observe that the MCAB tends to perform better in the cases with more MB-SL neighboring BSSs. In those situations, the $\overline{s}$ when using MCAB is above 94\% in half of the scenarios, whereas below 85\% when using the SLCI and MCAA. Although the optimal solution will be to avoid coexistence issues by not having any legacy BSSs, the periodic channel evaluation of the MCAB adds the required flexibility to minimize negative coexistence effects.

At last, Figure~\ref{fig:Coex_MBSL} shows the avg. satisfaction when $\text{BSS}_\text{A}$ is set as a legacy MB-SL with the aim to observe if the presence of MLO devices will benefit legacy ones. As previously, we incremented the fraction of MLO BSSs from 0, to 0.3, 0.7, and 1. Figure~\ref{fig:Coex_SLCI} reveals that legacy MB-SL BSSs can benefit from the fact of having MLO BSSs around them, as the improvement is highly noticeable. In fact, we observe that between the best and worst cases the satisfaction increases by a 40\% for half of the scenarios evaluated. Then, from the perspective of a legacy BSS, the adoption of the MLO by other BSSs represents also a performance improvement.

% ---------------

\section{Conclusions and future work}

In this letter, we assessed the implementation of a traffic manager to perform traffic allocation on top of MLO-capable BSSs. We evaluated three policy schemes under different conditions to shed some light on the potential performance gains of dynamic policies in comparison to non-dynamic ones. Under a wide variety of scenarios, our results shown that dynamic policies should be applied in presence of long-lasting flows, since their frequent adaptation to the instantaneous congestion conditions allows to minimize the effect of the neighboring AP~MLDs' actions. By the nature of video flows, it has been found also that the MCAB is able to maximize the traffic delivery by keeping a satisfaction ratio of 95\% for most of the evaluated scenarios. Under coexistence conditions, we observe that an excessive number of legacy BSSs may harm the performance of MLO ones. However, we found that the MCAB is able to reduce the negative impact of legacy BSSs by almost 10\% compared to MCAA, as it is able to react to changes in the channel occupancy of the different interfaces.

Regarding future research, we plan to extend current traffic management policies to also support link aggregation at channel access. Regarding improving QoS provisioning in next generation Wi-Fi networks, traffic differentiation policies should be further investigated in presence of heterogeneous stations, providing solutions that go beyond the default TID-to-link mapping functionality. Finally, we also consider the redesign of the traffic management module as part of an end-to-end Software Defined Networking solution, closely working with an external controller in charge of multiple APs to properly allocate traffic flows to interfaces.

\bibliographystyle{unsrt}
\bibliography{main}

\end{document}